# Deep Remix: Remixing Musical Mixtures Using a Convolutional Deep Neural Network

Andrew J.R. Simpson [#1], Gerard Roma [#2], Mark D. Plumbley [#3]

[#] Centre for Vision, Speech and Signal Processing, University of Surrey
Guildford, UK

[1] andrew.simpson@surrey.ac.uk
[2] g.roma@surrey.ac.uk
[3] m.plumbley@surrey.ac.uk

*Abstract*—**Audio source separation is a difficult machine learning problem and performance is measured by comparing extracted signals with the component source signals. However, if separation is motivated by the ultimate goal of *re-mixing* then complete separation is not necessary and hence separation difficulty and separation quality are dependent on the nature of the re-mix. Here, we use a convolutional deep neural network (DNN), trained to estimate 'ideal' binary masks for separating voice from music, to perform re-mixing of the vocal balance by operating directly on the individual magnitude components of the musical mixture spectrogram. Our results demonstrate that small changes in vocal gain may be applied with very little distortion to the ultimate re-mix. Our method may be useful for re-mixing existing mixes.**

*Index terms*—**Deep learning, supervised learning, convolution, source separation, audio re-mixing.**

## I. INTRODUCTION

Mixing is the process by which component sounds, known as sources, are superposed to form a mixture. In traditional acoustic music, mixing occurs in the acoustic domain and the balance is set by the musicians. In the electronic or digital domains, the balance between musical instruments is set by recording or mixing engineers. In both cases, the balance of the mix is critical to the aesthetic and perceptual qualities of the music [1]. Once a mix has been made, there are many situations in which it might be useful to adjust, or *re-mix*, the balance between the sources. However, like the mixing of ink and water, the deceptively simple process of mixing features an increase in entropy that is exquisitely difficult to undo.

Audio source separation addresses the problem of un-mixing and is often motivated by idealistic goals of complete separation. As a result, separation performance is quantified in terms of a comparison between the original sources and the extracted estimates [2]-[11]. However, if the ultimate goal of separation is to re-balance the mixture [1] then the source-focused separation quality measures may not be appropriate because they do not capture the ultimate quality of the re-mix.

A common approach to audio source separation is to apply a binary mask to a complex mixture spectrogram, where the mask identifies elements of the spectrogram that belong to a given source. A consequence of this approach is the production of a masked spectrogram featuring discontinuities that contribute to degraded sound quality that is measured using the source-focused separation quality measures [11]. However, if the goal of separation is to re-mix by applying new gains to the respective sources at the level of the individual components of the mixture spectrogram, the severity of the mask discontinuities is dependent on the gains applied. Hence, it is possible to parameterize the problem of separation-for-re-mix in terms of a trade-off between the magnitude of the changes made during re-mixing and the sound quality of the re-mix itself.

In a previous paper, we used multi-track recordings ('stems') to train a convolutional deep neural network (DNN) to make probabilstic predictions [4] of binary masks for extracting vocals from musical mixtures [5]. Here, using binary masks estimated using the voice-extraction convolutional DNN [5], we applied various gains to the vocal components within musical mixtures in order to simulate various separate-for-re-mix scenarios. We then compared these estimated re-mixes with idealized re-mixes (taken from the respective re-mixes of the original source components) using the BSS-Eval toolbox [11]. Our results show that re-mix quality is inversely proportional to the degree of re-mixing that is applied. In particular, our results demonstrate that there is a region of re-mix parameter space where objective measures of quality are good for small changes in gain.

## II. METHOD

We consider the problem of re-mixing a mixture within a simulated ensemble musical performance scenario that features a relatively wide variety of musical contexts, each with at least one vocal component. In each context, which we call 'a song', there are several musical 'accompaniment' signals. After the various signals are mixed together, we refer to the resulting mixture as 'a mix'. We used 63 fully produced and studio-recorded songs in the training and testing of the DNN [5]. The songs were taken from the MedleyDB database [12]; The average song duration was 3.7 minutes (standard deviation (STD): ±2.7 mins), the average number of accompaniment stems was 7.2 (STD: ±6.6 sources) and the average number of vocal tracks was 1.8 (STD: ±0.8 sources).

During the training and later testing of the DNN, for each song, the source signals were classified as either vocal or non-vocal (according to the labels assigned by the music producers). These sounds included male and female singing voice and 'rap'. Non-vocal sounds were typical accompanying instruments (drums, bass, guitars, piano, percussion, etc). Because the various source sounds were studio recorded they featured minimal bleed (interference) from other sources.

To produce the mixing contexts, source sounds were peak normalized before being linearly summed into either a vocal mixture or a non-vocal mixture respectively. Then, the separate vocal and non-vocal mixtures were each peak normalized and summed to provide a complete mix. This provided a mixture for each song, similar to that of a human mixing engineer, which served as both training (for separation - see [5]) and also served as test mixtures for later re-mixing.

All signals and mixtures were monaural and were sampled at a rate of 44.1 kHz. For training, the respective source (vocal / non-vocal) and mixture signals were transformed into spectrograms using the short-time Fourier transform (STFT) with window size of 2048 samples, overlap interval of 512 samples and a Hanning window, giving spectrograms with 1025 frequency bins. From the source spectrograms a binary mask was computed and the DNN was trained on these ideal binary masks for the first 50 songs of the dataset. The magnitude-only mixture spectrograms computed from the first 50 songs and the respective ideal binary masks were used as training data. This left 13 songs for use as test data. Note, phase was not used in training the model.

Summarising in brief the method outlined in [5]; The mixture spectrograms and the corresponding source spectrograms were cut up into corresponding windows of 20 samples (in time). The windows shifted at intervals of 60 samples (i.e., there was no overlap). Thus, for every 20-sample window, for training the models there was a mixture spectrogram matrix of size 1025x20 (frequency bins x time) samples and an ideal binary mask matrix of the same size. This gave approximately 15,000 training examples. The spectrograms for the 13 test songs were cut up with overlap intervals of 1 sample.

We used a feed-forward DNN of size 20500x20500x20500 units (1025 x 20 = 20500) featuring the biased-sigmoid activation function [13] throughout with zero bias for the output layer. Each spectrogram window of size 1025 x 20 was unpacked into a vector of length 20500. The input layer was the mixture spectrogram (20500 samples) and the DNN was trained to synthesize the ideal binary mask at its output layer. 100 full iterations of stochastic gradient descent (SGD) were used for training, each iteration (of SGD) comprising a full sweep of the training data.

*Probabilistic Binary Mask*. Using the DNN to make predictions in a sliding window manner, overlapping at 1-sample intervals, we obtained a distribution (size 20) of mask predictions for each time-frequency element of the mixture spectrogram [4], [5]. We then took the mean of these predictions and subjected it to a 'confidence' threshold ($\alpha$);

$$M^V_{t,f} = \begin{cases} 1 & for \quad \frac{1}{T}\sum_{i=0}^{T} S_{t+i,f} > \alpha \\ 0 & for \quad \frac{1}{T}\sum_{i=0}^{T} S_{t+i,f} \leq \alpha \end{cases} \quad (1)$$

where $M^V$ refers to the binary mask for the vocal source, $T$ refers to the window size (20), $t$ is the time index, $i$ is the window index and $f$ is the frequency (bin) index into the estimated mask ($S$). This allowed us to adjust $\alpha$ to produce vocal masks at different levels of confidence. The vocal mask was then converted from a binary mask to a scaling matrix ($Z^V$) by first multiplying non-zero elements with a gain ($g$) and then assigning the value 1 to all zero elements;

$$Z^V_{t,f} = \begin{cases} 1 & for \quad M^V_{t,f} = 0 \\ gM^V_{t,f} & for \quad M^V_{t,f} = 1 \end{cases} \quad (2)$$

This meant that when the scalar mask was multiplied with the complex spectrogram, the vocal components were changed in gain whilst the background/accompanying instruments were unaffected. The resulting re-mixed spectrograms were inverted with a standard overlap-and-add procedure. For reference, the same re-mixes were also produced by applying the gain ($g$) to the original vocal signal before summing with the un-altered non-vocal signal. Re-mix quality was then measured using the BSS-EVAL toolbox [11] and is quantified in terms of signal-to-artefact ratio (SAR) computed by comparing the estimated re-mix with the reference (linear) re-mix. I.e., separation quality was not directly measured and the separate component sources were not taken into account in the quality measurement. Re-mix quality was assessed at different confidence levels by setting different values of $\alpha$ and in different re-mix contexts by setting different values of $g$. Note that, although $g$ is defined as a scalar, we refer to its value in decibels [gain in dB = $10log10(g)$] from here onwards.

## III. RESULTS

Figure 1 plots surface contours representing mean SAR (across the 13 test songs) as a function of both gain ($g$) applied to the vocal component of the masked mixture and confidence ($\alpha$) used in constructing the respective masks. The part of the plot where $\alpha = 0$ represents a baseline because it captures the distortion (SAR) observed when the unprocessed starting mixture is compared with the respective ideal re-mix. Since this ratio tends towards infinity as the starting mixture and re-mix converge (at $g = 0$ dB), there is little interpretable information to capture near this point. Therefore, the plots feature the range of 5>$g$>20 dB, where there is a meaningful difference between the measurable effect of the masked manipulation and the baseline. Furthermore, the plots have been split into negative gains ($g$<-5 dB) and positive gains ($g$>5 dB) so that their ranges might be optimized for the visualization. For both negative (Fig. 1a) and positive (Fig. 2a) gains distortion (SAR) worsens with increasing gain and both feature a maximum around $α = 0.3$. Note that, taking into account the different scaling on the z-axes of the two respective plots, there is a marked asymmetry between the functions. This is the result of the energy imbalance between two respective signals (which causes the ratio function to be centered at a non-zero location).

Fig. 2 plots cross-sections (with 95% confidence intervals) from the same data as a function of gain ($g$) for $α = 0.1$ (Fig. 2b), $α = 0.3$ (Fig. 2c), $α = 0.6$ (Fig. 2d) and $α = 0.9$ (Fig. 2e) respectively. Fig. 2a also plots the baseline for reference, which is also superposed on the various plots (Figs. 2b-e). Note again the asymmetry (even in the baseline function, which is equivalent to $α = 0$ in Fig. 1) that results from energy imbalance as noted above for Fig. 1. At small values of $α$ ($α = 0.1$-$0.6$; Figs. 2b-d), there is a measurable degree of successful re-mixing in evidence (SAR > baseline) but at $α = 0.9$ (Fig. 2e) performance is at or below baseline. This is interesting given that $α$ represents confidence in the mask predictions of the DNN (large values of $α$ were most successful in reliably separating voice from background in [5]). Generally, these results illustrate the trade-off between certainty (parameterized by $α$) in correcty adjusting the gain of vocal components and the fact that $α$ is an index into the cumulative distribution function, meaning that when we are very certain of the classification we have very little signal to adjust. Furthermore, since there is no concept of 'interference' in this re-mix paradigm, SAR captures a combination of distortion factors including incorrect mixture ratios and nonlinear distortion products. Therefore, the peak in the function represents a cross-over point between correct adjustment of vocal components, incorrect adjustment of accompaniment and any related distortion products.

In order to illustrate the orthogonal cross section of the surface plots of Fig. 1, Figure 3 plots mean SAR as a function of $α$ for gains of -20 and 20 dB respectively (shaded areas represent 95% confidence intervals). This plot illustrates more clearly the mutual peak around $α = 0.3$.

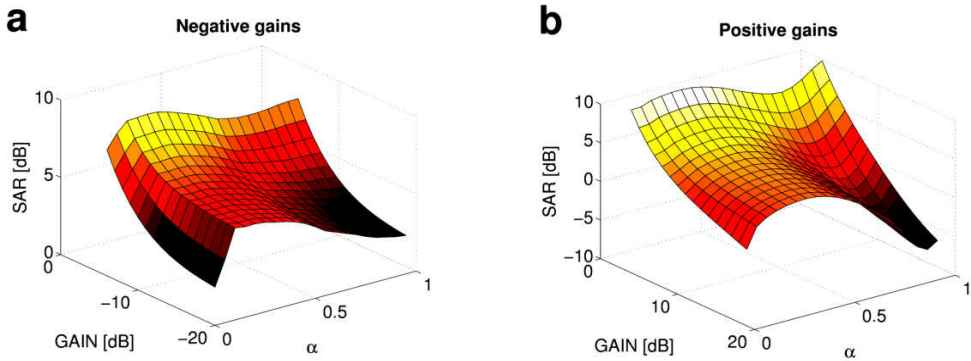

**Fig. 1. Re-mixing vocals using a probabilistic convolutional DNN: Mean re-mix distortion versus vocal gain and $α$.** The plots of this figure represent the overall re-mix distortion (mean SAR across the 13 test songs) for various vocal gains and various levels of confidence. **a** plots negative gains (i.e., vocals are quieter) and **b** plots positive gains (i.e., vocals are louder). Note the z-axis (SAR) is not the same for both plots.

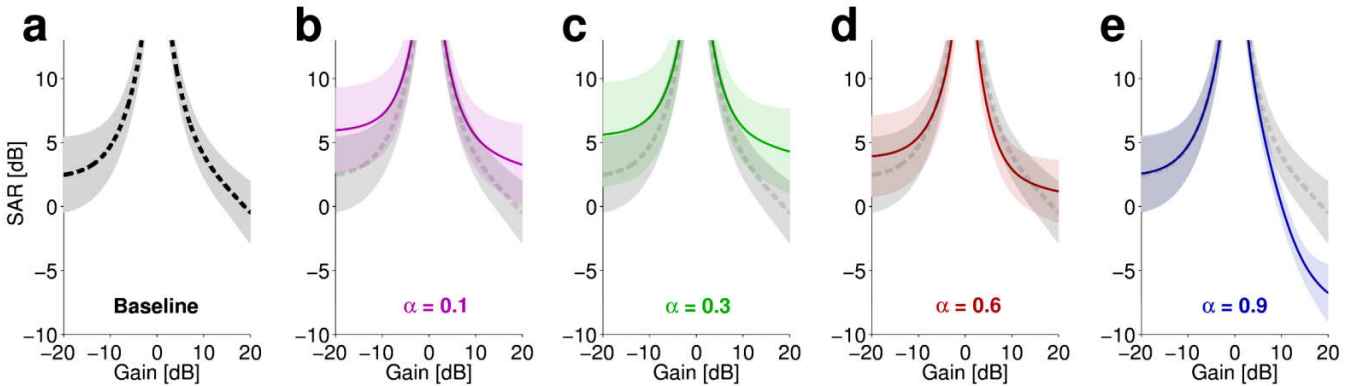

**Fig. 2. Re-mix distortion as a function of gain ($g$): a** plots a baseline comparing the starting mixes and the ideal re-mixes. Panels **a** to **e** plot plot SAR for the re-mixes computed using the probabilistic convolutionanl DNN (as compared with the ideal re-mixes) for $α = 0.1$ (**b**), 0.3 (**c**), 0.6 (**d**) and 0.9 (**e**) respectively. Baseline is included on all plots for reference. Lines represent mean (across the 13 test songs) and shaded areas represent 95% confidence intervals.

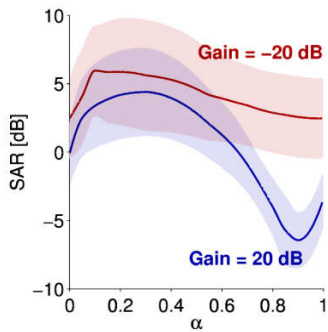

**Fig. 3. Re-mix distortion as a function of mask confidence (α):** Mean SAR (across the 13 test songs) for the re-mixes computed using the convolutionanl DNN (as compared with the ideal re-mixes) for $g$ = -20 dB (red) and 20 dB (blue) respectively. Shaded areas represent 95% confidence intervals.

## IV. DISCUSSION AND CONCLUSION

We have demonstrated that a convolutional deep neural network, trained to separate vocal sounds from within musical mixtures, can be used to re-mix the vocal balance in new mixtures. We have also illustrated the trade-off between re-mix quality and the scale of the gain change introduced for a given re-mix. This means that very small adjustments to the mix can be made for relatively little cost in terms of distortion.

Our paradigm features separation/re-mixing which starts from arbitrary mixtures, constructed by peak normalization (of both sub-stems and vocal-/non-vocal stems). Although these mixtures crudely resemble the kind of mixing that human engineers might do, the present results tend to suggest that the starting point for the re-mix may be critical. In particular, because the approach is based on STFT decomposition, and because it is based on the magnitude component only, the starting gains are critical. Therefore, future work is necessary to generalize the present paradigm to more realistic mixing scenarios, where the target mixtures represent some prior attempt at mixing (presumably by a human). In addition, we have considered relatively crude re-mixing, featuring large/coarse gain changes that may not represent the needs of a typical re-mix problem. Therefore, it remains to be seen how the present approach will generalize to more realistic re-mix scenarios and to sources other than vocals. We also note that our convolutional DNN was trained with relatively little data (and relatively few iterations of SGD) and performance can likely be improved substantially by improvement of the training regime.

## ACKNOWLEDGMENT

AJRS, GR and MDP were supported by grant EP/L027119/1 from the UK Engineering and Physical Sciences Research Council (EPSRC). Data and materials are available from the authors on request.